\documentclass[twocolumn,aps,prc,superscriptaddress,showpacs,floatfix]{revtex4}
\usepackage{amssymb}
\usepackage{graphicx}

\begin{document}

\title{Anisotropic flow in Cu+Au collisions at $\sqrt{s_{NN}}=200$ GeV}
\author{Lie-Wen Chen}
\affiliation{Institute of Theoretical Physics, Shanghai Jiao Tong University, Shanghai
200240, China}
\affiliation{Center of Theoretical Nuclear Physics, National Laboratory of Heavy Ion
Accelerator, Lanzhou 730000, China}
\author{Che Ming Ko}
\affiliation{Cyclotron Institute and Physics Department, Texas A\&M University, College
Station, Texas 77843-3366}
\date{\today }

\begin{abstract}
The anisotropic flow of charged hadrons in asymmetric Cu+Au collisions at
the Relativistic Heavy Ion Collider is studied in a multi-phase transport
model. Compared with previous results for symmetric Au+Au collisions,
charged hadrons produced around midrapidity in asymmetric collisions are
found to have a stronger directed flow $v_{1}$ and their elliptic flow $%
v_{2} $ is also more sensitive to the parton scattering cross section. While
higher-order flows $v_{3}$ and $v_{4}$ are small at all rapidities, both $%
v_{1}$ and $v_{2}$ in these collisions are appreciable and show an asymmetry
in forward and backward rapidities.
\end{abstract}

\maketitle

\section{Introduction}

There have been extensive studies on the azimuthal anisotropy of hadron
momentum distributions in the transverse plane perpendicular to the beam
direction, particularly the elliptic flow $v_{2}$, in heavy-ion collisions
at various energies \cite{reisdorf}. The hadron transverse momentum
anisotropy is generated by the pressure anisotropy in the initial compressed
matter formed in non-central heavy ion collisions \cite{Barrette94, Appel98}
and is sensitive to the properties of produced matter in these collisions.
For heavy-ion collisions at the Relativistic Heavy Ion Collider (\textrm{RHIC%
}), it has been shown that this sensitivity not only exists in the larger
elliptic flow \cite%
{Ollit92,Rqmd,Danie98,Zhang99,Zheng99,gyulv2,Lin:2001zk,Voloshin03} but also
in smaller higher-order anisotropic flows \cite%
{Kolb99,Teaney99,Kolb00,Kolb03,STAR04,chen04,Kolb04}. To investigate the
influence due to initial collision geometry on anisotropic flows in heavy
ion collisions, one usually varies the impact parameter of a collision or
the atomic number of colliding nuclei \cite{cucu05}. Another possibility is
to study asymmetric heavy ion collisions with unequal mass nuclei as their
initial collision geometry is different from that in collisions with equal
mass nuclei.

In the present work, we use a multi-phase transport (\textrm{AMPT}) model,
that includes both initial partonic and final hadronic interactions \cite%
{Zhang:2000bd,Lin:2001cx}, to study the anisotropic flows $v_{1}$, $v_{2}$, $%
v_{3}$, and $v_{4}$ of charged hadrons in asymmetric Cu+Au collisions at $%
\sqrt{s}=200$ \textrm{AGeV} at \textrm{RHIC}. Use is made of both the
default version and the version with string melting, i.e., allowing hadrons
that are expected to be formed from initial strings to convert to their
valence quarks and antiquarks \cite{Lin:2001zk,LinHBT02,ko}. The latter was
able to explain measured $p_{T}$ dependence of $v_{2}$ and $v_{4}$ of
mid-rapidity charged hadrons with a parton scattering cross section of about 
$10$ \textrm{mb} while the default version seemed to give a better
description of the anisotropic flows at large rapidities where the string
degree of freedom dominates \cite{v13plb05}. We find that charged hadrons
produced around midrapidity in asymmetric Cu+Au collisions have a stronger
directed flow $v_{1}$ and their elliptic flow is also more sensitive to the
parton scattering cross section than those found previously in symmetric
Au+Au collisions. Furthermore, both $v_{1}$ and $v_{2}$ of charged hadrons
are asymmetric in the forward and backward rapidities. The higher-order
flows $v_{3}$ and $v_{4}$ are generally small, and their sensitivity to the
parton scattering cross section is thus less clear.

This paper is organized as follows. In Sec. \ref{ampt}, the AMPT model is
briefly reviewed. Results on the pseudorapidity dependence of the
anisotropic flows of charged hadrons are shown in Sec. \ref{eta}. The $p_{T}$
dependence of the anisotropic flows of charged hadrons around midrapidity is
studied in Sec. \ref{mideta} while that at large rapidities is shown in Sec. %
\ref{largeeta}. Finally, a brief summary is given in Sec. \ref{summary}.

\section{The AMPT model}

\label{ampt}

The \textrm{AMPT} model \cite{Zhang:2000bd,Lin:2001cx,zhang,pal,ampt} is a
hybrid model that uses minijet partons from hard processes and strings from
soft processes in the Heavy Ion Jet Interaction Generator (\textrm{HIJING})
model \cite{Wang:1991ht} as the initial conditions for modeling heavy ion
collisions at ultra-relativistic energies. Time evolution of resulting
minijet partons is described by Zhang's parton cascade (\textrm{ZPC}) \cite%
{Zhang:1997ej} model. At present, this model includes only parton-parton
elastic scatterings with an in-medium cross section given by: 
\begin{equation}
\frac{d\sigma _{p}}{dt}\approx \frac{9\pi \alpha _{s}^{2}}{2}\left( 1+{\frac{%
{\mu ^{2}}}{s}}\right) \frac{1}{(t-\mu ^{2})^{2}},  \label{crscp}
\end{equation}%
where the strong coupling constant $\alpha _{s}$ is taken to be $0.47$, and $%
s$ and $t$ are usual Mandelstam variables. The effective screening mass $\mu 
$ depends on the temperature and density of the partonic matter but is taken
as a parameter in \textrm{ZPC} for fixing the magnitude and angular
distribution of the parton scattering cross section. After minijet partons
stop interacting, they are combined with their parent strings, as in the 
\textrm{HIJING} model with jet quenching, to fragment into hadrons using the
Lund string fragmentation model as implemented in the \textrm{PYTHIA}
program \cite{Sjostrand:1994yb}. Final-state scatterings among hadrons are
then modelled by a relativistic transport (\textrm{ART}) model \cite%
{Li:1995pr}. The default \textrm{AMPT} model \cite{Zhang:2000bd} has been
quite successful in describing measured rapidity distributions of charge
particles, particle to antiparticle ratios, and spectra of low transverse
momentum pions and kaons \cite{Lin:2001cx} in heavy ion collisions at the
Super Proton Synchrotron (\textrm{SPS}) and \textrm{RHIC}. It has also been
useful in understanding the production of $J/\psi $ \cite{zhang} and
multistrange baryons \cite{pal} in these collisions.

Since the initial energy density in heavy ion collisions at \textrm{RHIC} is
much larger than the critical energy density at which the hadronic matter to
quark-gluon plasma transition would occur \cite{Kharzeev:2001ph,zhang,cucu05}%
, the \textrm{AMPT} model has been extended to convert initial excited
strings into partons \cite{Lin:2001zk}. In this string melting scenario,
hadrons (mostly pions), that would have been produced from string
fragmentation, are converted instead to valence quarks and/or antiquarks
with current quark masses. Interactions among these partons are again
described by the\textrm{\ ZPC} parton cascade model. Since inelastic
scatterings are not included, only quarks and antiquarks from melted strings
are present in the partonic matter. The transition from the partonic matter
to the hadronic matter is achieved using a simple coalescence model, which
combines two nearest quark and antiquark into mesons and three nearest
quarks or antiquarks into baryons or anti-baryons that are close to the
invariant mass of these partons. The present coalescence model is thus
somewhat different from the ones recently used extensively \cite%
{greco,hwa,fries,molnar03} for studying hadron production at intermediate
transverse momenta. Using parton scattering cross sections of $6$-$10$ 
\textrm{mb}, the \textrm{AMPT} model with string melting was able to
reproduce both the centrality and transverse momentum (below $2$ \textrm{GeV}%
$/c$) dependence of the elliptic flow \cite{Lin:2001zk} and pion
interferometry \cite{LinHBT02} measured in Au+Au collisions at $\sqrt{s}=130$
\textrm{AGeV} at \textrm{RHIC} \cite{Ackermann:2000tr,STARhbt01}. It has
also been used for studying the kaon interferometry \cite{lin} and charm
flow \cite{zhangc} in these collisions. We note that above parton cross
sections are significantly smaller than that needed to reproduce the parton
elliptic flow from the hydrodynamic model \cite{molnar}. The resulting
hadron elliptic flows in the \textrm{AMPT} model with string melting are,
however, amplified by modelling hadronization via quark coalescence \cite%
{molnar03}, leading to a satisfactory reproduction of experimental data. As
shown earlier in Refs. \cite{csorgo,anderlik}, the hadronization of the
quark-gluon plasma leads in general to such acceleration at freeze-out.

\section{Pseudorapidity dependence of anisotropic flows}

\label{eta}

The anisotropic flows $v_{n}$ of particles in heavy ion collisions are the
Fourier coefficients in the decomposition of their transverse momentum
spectra in the azimuthal angle $\phi $ with respect to the reaction plane 
\cite{Posk98}, i.e., 
\begin{equation}
E\frac{d^{3}N}{dp^{3}}=\frac{1}{2\pi }\frac{dN}{p_{T}dp_{T}dy}\left[
1+\sum_{n=1}^{\infty }2v_{n}(p_{T},y)\cos (n\phi )\right]  \label{dndphi}
\end{equation}%
Because of the symmetry $\phi \leftrightarrow -\phi $ in the collision
geometry, sine terms do not appear in above expansion. Since the projectile
and target sides of the reaction plane can in principle be identified
experimentally, all isotropic flows $v_{n}$'s are generally finite. Although
in collisions with equal mass nuclei the odd-order anisotropic flows of
particles vanish as a result of the symmetry $(y, \phi)\to(-y, \phi+\pi)$,
at midrapidity and midrapidity only, in case of collisions of asymmetric
nuclei the odd flow components are finite at midrapidity also.

Anisotropic flows depend on both particle transverse momentum and rapidity,
and for a given rapidity the anisotropic flows at transverse momentum $p_{T}$
can be evaluated according to 
\begin{equation}
v_{n}(p_{T})=\left\langle \cos (n\phi )\right\rangle ,  \label{vn1}
\end{equation}%
where $\left\langle \cdot \cdot \cdot \right\rangle $ denotes average over
the azimuthal distribution of particles with transverse momentum $p_{T}$.

The anisotropic flows $v_{n}$ can further be expressed in terms of
single-particle averages: 
\begin{eqnarray}
v_{1}(p_{T}) &=&\left\langle \frac{p_{x}}{p_{T}}\right\rangle  \label{v1} \\
v_{2}(p_{T}) &=&\left\langle \frac{p_{x}^{2}-p_{y}^{2}}{p_{T}^{2}}%
\right\rangle  \label{v2} \\
v_{3}(p_{T}) &=&\left\langle \frac{p_{x}^{3}-3p_{x}p_{y}^{2}}{p_{T}^{3}}%
\right\rangle  \label{v3} \\
v_{4}(p_{T}) &=&\left\langle \frac{p_{x}^{4}-6p_{x}^{2}p_{y}^{2}+p_{y}^{4}}{%
p_{T}^{4}}\right\rangle  \label{v4}
\end{eqnarray}%
where $p_{x}$ and $p_{y}$ are, respectively, projections of particle
momentum in and perpendicular to the reaction plane. Since the \textrm{AMPT}
model also provides information on the spatial anisotropy of colliding
nuclear matter, which is responsible for generating the momentum anisotropic
flows, it is of interest to introduce the spatial anisotropic coefficient $%
s_{n}$ by expressions similar to those for the anisotropic flows $v_{n}$ but
in terms of the spatial distributions of particles in the transverse plane.

\begin{figure}[th]
\includegraphics[width=3.4in,height=3.5in]{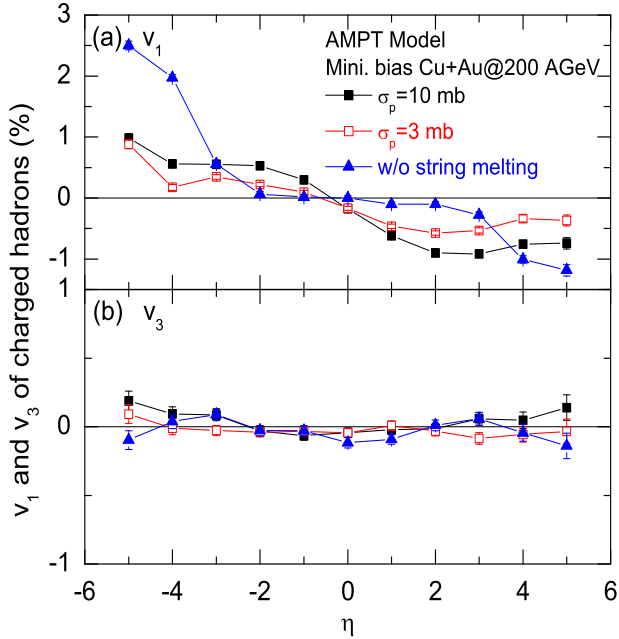}
\caption{{\protect\small (Color online) Pseudorapidity dependence of }$v_{1}$%
{\protect\small \ (a) and }$v_{3}${\protect\small \ (b) for charged hadrons
from minimum bias events of Cu+Au collisions at }$\protect\sqrt{s}=200$%
{\protect\small \ AGeV by using the string melting scenario with parton
scattering cross sections }$\protect\sigma _{p}=3${\protect\small \ (open
squares) and }$10${\protect\small \ mb (solid squares) and the default AMPT
model without string melting (solid triangles).}}
\label{v13eta}
\end{figure}

In Figs. \ref{v13eta} (a) and \ref{v13eta}(b), we show the pseudorapidity
dependence of $v_{1}$ and $v_{3}$, respectively, for charged hadrons from
minimum bias events of Cu+Au collisions at $\sqrt{s}=200$ \textrm{AGeV} by
using the string melting scenario with parton scattering cross sections $%
\sigma _{p}=3$ (open squares) and $10$ \textrm{mb} (solid squares) and also
the scenario without string melting (default \textrm{AMPT} model, solid
triangles). Compared with the AMPT results shown in Fig. 2 of Ref. \cite%
{v13plb05} for Au+Au collisions at same c.m. energy per nucleon, where the
string melting scenario with a parton scattering cross section of $10$ 
\textrm{mb} describes very well the experimental data around midrapidity
from the STAR collaboration \cite{STAR04}, asymmetric Cu+Au collisions
display clearly a stronger $v_{1}$ around mid-$\eta $ in the AMPT model with
string melting. Also, the pseudorapidity dependence of $v_{1}$ has a
negative slope around mid-$\eta $ and is sensitive to the parton scattering
cross section. It is further seen from Fig. \ref{v13eta} (a) that the $p_{T}$%
-integrated $v_{1}$ is very small around mid-$\eta $ in the default AMPT
model without string melting. In addition, $v_{1}$ displays an asymmetry in
forward and backward rapidities around mid-$\eta $ with the magnitude of $%
v_{1}$ at forward rapidity larger than that at backward rapidity. As shown
later, this asymmetry is related to the different spatial deformation at
forward and backward rapidities around mid-$\eta $. For the scenario without
string melting, an even stronger asymmetry of $v_{1}$ in forward and
backward rapidities is seen at larger pseudorapidity ($\left\vert \eta
\right\vert \geq 3$), with the magnitude of $v_{1}$ at large backward
pseudorapidity (Au-like rapidity) about a factor of $2$ larger than that at
large forward pseudorapidity (Cu-like rapidity). This result can be
understood from the fact that hadronic degrees of freedom dominate at large
pseudorapidities as the scenario without string melting has been shown in
Ref.\cite{v13plb05} to give a better description of the STAR data \cite%
{STAR04}, and there are thus more hadrons and hadronic rescattering at the
Au-like rapidity. We will return to this point later. On the other hand, the
magnitude of $p_{T}$-integrated $v_{3}$ is very small (less than $0.3\%$) in
the whole pseudorapidity region for all three scenarios considered here.

\begin{figure}[th]
\includegraphics[scale=0.95]{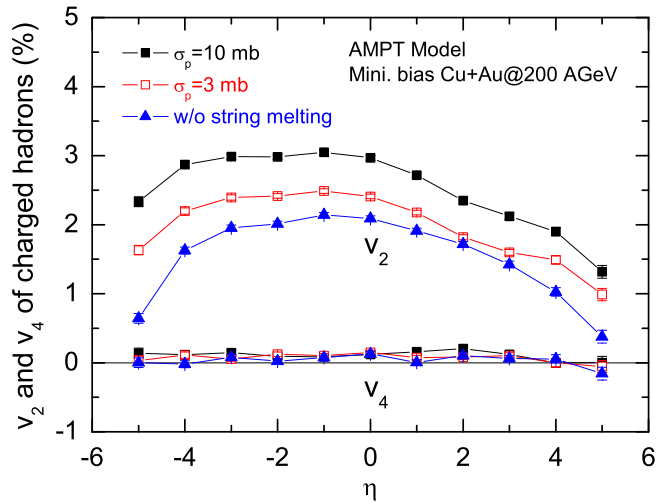}
\caption{{\protect\small (Color online) Same as Fig. \protect\ref{v13eta}
for }$v_{2} ${\protect\small \ and }$v_{4}${\protect\small .}}
\label{v24eta}
\end{figure}

The pseudorapidity dependence of $v_{2}$ and $v_{4}$ for charged hadrons
from minimum bias events of Cu+Au collisions at $\sqrt{s}=200$ \textrm{AGeV}
is shown in Fig. \ref{v24eta} by using same three scenarios as in Fig. \ref%
{v13eta}. Similar to $v_{1}$ shown in Fig. \ref{v13eta}, the elliptic flow $%
v_{2}$ displays a clear asymmetry in forward and backward rapidities around
mid-$\eta $ as well as at large pseudorapidity. In the case of string
melting scenario with a parton scattering cross section of $10$ \textrm{mb},
the value of $v_{2}$ around mid-$\eta $\ is about $3\%$. Compared with the $%
v_{2}$ value of about $4.5\%$ around mid-$\eta $\ in Au+Au collisions at
same c.m. energy per nucleon shown in Fig. 3 of Ref. \cite{v13plb05}, the
scaling of elliptic flow according to the reaction system size as proposed
in Ref. \cite{cucu05} is also satisfied in asymmetric collisions. As in the
case of $v_{3}$ shown in Fig. \ref{v13eta} (b), the magnitude of
higher-order $p_{T}$-integrated $v_{4}$ is also very small (less than $0.3\%$%
) in the whole pseudorapidity region for all three scenarios.

\section{$p_{T}$-dependence of anisotropic flows around midrapidity}

\label{mideta}

\begin{figure}[th]
\includegraphics[scale=1.2]{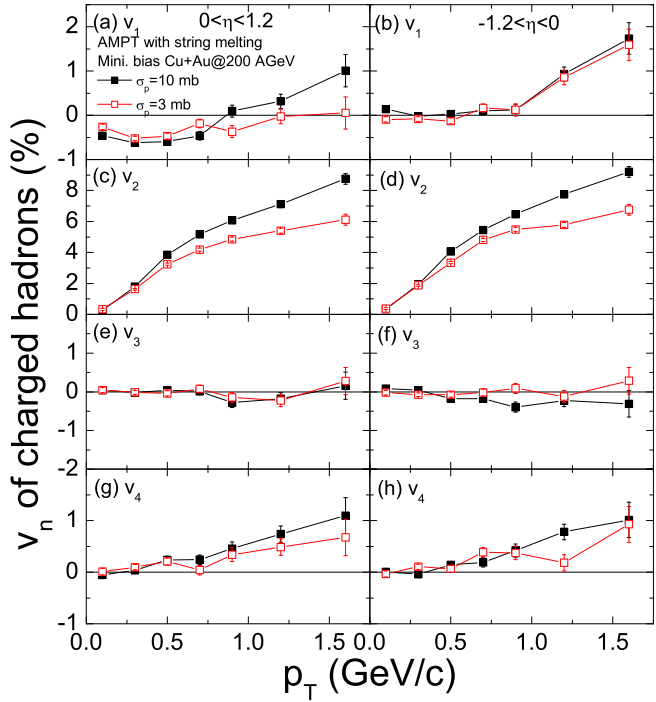}
\caption{{\protect\small (Color online) }$p_{T}${\protect\small -dependence
of }$v_{1}${\protect\small \ ((a) and (b)) }$v_{2}${\protect\small \ ((c)
and (d)), }$v_{3}${\protect\small \ ((e) and (f)), and }$v_{4}$%
{\protect\small \ ((g) and (h)) for charged hadrons at forward mid-}$\protect%
\eta ${\protect\small \ (}$0<\protect\eta <1.2${\protect\small , left
panels) and backward mid-}$\protect\eta ${\protect\small \ (}$-1.2<\protect%
\eta <0${\protect\small , right panels) from minimum bias events of Cu+Au
collisions at }$\protect\sqrt{s}=200${\protect\small \ AGeV.}}
\label{vnPTeta12}
\end{figure}

More detailed information on anisotropic flows can be obtained from the
differential anisotropic flows, i.e., their $p_{T}$ dependence. For hadrons
produced around mid-$\eta $ in heavy ion collisions at \textrm{RHIC}, the
partonic degree of freedom is expected to be important and the scenario with
string melting describes very well the observed anisotropic flows of charged
hadrons \cite{v13plb05}. In Fig. \ref{vnPTeta12}, we show results obtained
with the string melting scenario on the $p_{T}$-dependence of $v_{1}$, $%
v_{2} $, $v_{3}$, and $v_{4}$ for charged hadrons at forward mid-$\eta $ ($%
0<\eta <1.2$, left panels) and backward mid-$\eta $ ($-1.2<\eta <0$, right
panels) from minimum bias events of Cu+Au collisions at $\sqrt{s}=200 $ 
\textrm{AGeV}. Both parton scattering cross sections $\sigma _{p}=3$ (open
squares) and $10$ \textrm{mb} (solid squares) are used. The sensitivity to
the parton cross section is clearly seen in $v_{2}$ and $v_{4} $ at both
forward and backward mid-$\eta $. We note that the elliptic flow exhibits a
stronger sensitivity to the parton cross section in asymmetric Cu+Au
collisions than in symmetric Au+Au collisions. For example, the ratio of $%
v_{2}$ with $\sigma _{p}=10$ \textrm{mb} to that with $\sigma _{p}=3$ 
\textrm{mb} is about $1.4$ in Cu+Au collisions but is reduced to about $1.2$
in Au+Au collisions as given in Fig. 2 of Ref. \cite{chen04}. A similar
conclusion is obtained for $v_{4}$. Sensitivities of the $p_{T}$-dependence
of $v_{1}$ at forward mid-$\eta $ and $v_{3}$ at backward mid-$\eta $ to the
parton cross section are also seen.

Fig. \ref{vnPTeta12} also shows that the $p_{T}$-dependence of $v_{1}$ at
forward and backward mid-$\eta $ exhibits very different behaviors. At
forward mid-$\eta $, $v_{1}(p_{T})$ is non-zero and changes from negative to
positive values at a balance transverse momentum while $v_{1}(p_{T})$ at
backward mid-$\eta $ is essentially zero at low $p_{T}$ and becomes large
and positive above about $0.8$ \textrm{GeV}/\textsl{c} for both parton cross
sections. As a result, the $p_{T}$-integrated $v_{1}$ of charged hadrons
around mid-$\eta $ displays the asymmetry in forward and backward rapidities
shown in Figs. \ref{v13eta} (a). For $v_{2}$, its value at backward mid-$%
\eta $ is slightly larger than at forward mid-$\eta $ while the value of $%
v_{4}$ exhibits an opposite behavior. The difference between the $p_{T}$%
-dependence of $v_{3}$ at forward and backward mid-$\eta $ is, on the other
hand, not so clear due to their small values.

\begin{figure}[th]
\includegraphics[scale=0.9]{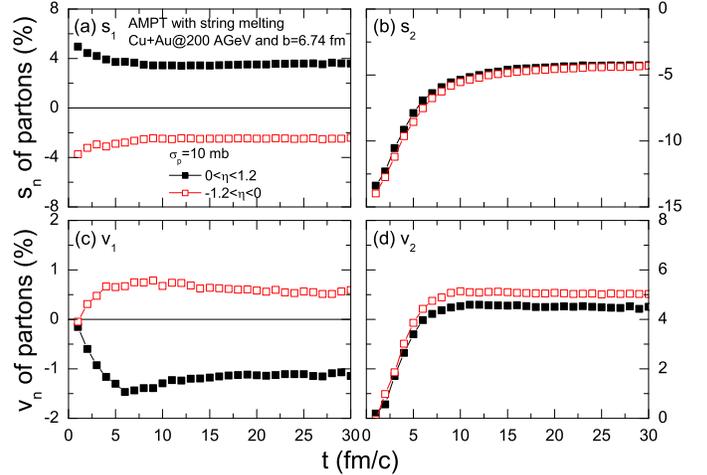}
\caption{{\protect\small (Color online) Time evolutions of }$s_{1}$%
{\protect\small \ (a), }$s_{2}${\protect\small \ (b), }$v_{1}$%
{\protect\small \ (c) and }$v_{2}${\protect\small \ (d) of partons at
forward mid-}$\protect\eta ${\protect\small \ (}$0<\protect\eta <1.2$%
{\protect\small , solid squares) and backward mid-}$\protect\eta $%
{\protect\small \ (}$-1.2<\protect\eta <0${\protect\small , open squares)
from Cu+Au collisions at }$\protect\sqrt{s}=200${\protect\small \ AGeV and }$%
b=6.74${\protect\small \ fm in the scenario of string melting with }$\protect%
\sigma _{p}=10${\protect\small \ mb.}}
\label{sv12Time}
\end{figure}

\begin{figure}[th]
\includegraphics[scale=0.85]{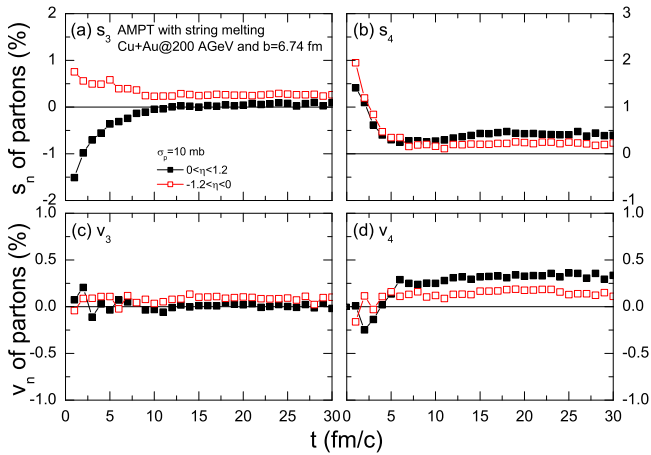}
\caption{{\protect\small (Color online) Similar as Fig. \protect\ref%
{sv12Time} for }$s_{3}${\protect\small \ (a), }$s_{4}${\protect\small \ (b), 
}$v_{3}${\protect\small \ (c) and }$v_{4}${\protect\small \ (d).}}
\label{sv34Time}
\end{figure}

The asymmetry in forward and backward rapidities around mid-$\eta $ observed
in Fig. \ref{vnPTeta12} can be better understood from the time evolutions of
parton spatial deformation and anisotropic flows as the latter are
transferred to those of hadrons when they are formed from quark and/or
antiquark coalescence. Also, scatterings among hadrons, which are included
in the \textrm{AMPT} model, do not affect much hadron anisotropic flows as a
result of small spatial anisotropy and low pressure in hadronic matter \cite%
{Lin:2001zk}. In Fig. \ref{sv12Time}, we show the time evolutions of $s_{1}$%
, $s_{2}$, $v_{1}$ and $v_{2}$ of partons at forward mid-$\eta $ ($0<\eta
<1.2$, solid squares) and backward mid-$\eta $ ($-1.2<\eta <0$, open
squares) from Cu+Au collisions at $\sqrt{s}=200$ \textrm{AGeV} and $b=6.74$
fm in the scenario of string melting with $\sigma _{p}=10$ \textrm{mb}. We
note that the impact parameter $b=6.74$ fm corresponds to the same $b/b_{%
\mathrm{max}}$ as in Au+Au collisions at $b=8$ fm, where $b_{\mathrm{max}}$
is the sum of the radii of colliding nuclei. Similar results for $s_{3}$, $%
s_{4}$, $v_{3}$ and $v_{4}$ are shown in Fig. \ref{sv34Time}.

It is seen from Fig. \ref{sv12Time} (a) that the strength of the spatial
deformation coefficient $s_{1}$ is initially large and decreases with time.
It reaches a saturation value at about $10$ \textrm{fm/}$c$ for partons at
both forward and backward mid-$\eta $. The strength of $s_{1}$ at forward
mid-$\eta $ is, however, larger than that at backward mid-$\eta $, leading
thus to a stronger $v_{1}$ at forward mid-$\eta $ as seen in Fig. \ref%
{sv12Time} (c) and also in Fig. \ref{v13eta} (a). On the other hand, Fig. %
\ref{sv12Time} (b) shows that the strength of parton spatial elliptic
deformation $s_{2}$ is slightly larger at backward mid-$\eta $ than at
forward mid-$\eta $, resulting in a slightly stronger $v_{2}$ at backward
mid-$\eta $ that is shown in Fig. \ref{sv12Time} (d) and also in Figs. \ref%
{v24eta} and \ref{vnPTeta12}. Therefore, the observed asymmetry of $v_{1}$
and $v_{2}$ in forward and backward rapidities around mid-$\eta $ is due to
an asymmetry in initial spatial deformation in asymmetric heavy ion
collisions.

From Fig. \ref{sv34Time} (a), we observe that the $s_{3}$ displays different
time evolutions for partons at forward and backward mid-$\eta $ and that its
saturated value at backward mid-$\eta $ is larger than at forward mid-$\eta $%
. As a result, a stronger $v_{3}$ at backward mid-$\eta $ as shown in Fig. %
\ref{sv34Time} (c) is seen. For $s_{4}$, it also displays different time
evolutions for partons at forward and backward mid-$\eta $. Its saturated
value is, on the other hand, larger at forward than backward mid-$\eta $,
leading to a stronger $v_{4}$ at forward mid-$\eta $ as shown in Fig. \ref%
{sv34Time} (d) and also in Fig. \ref{vnPTeta12}. However, because of its
small value, the asymmetry in forward and backward rapidities is not so
clear for the $v_{3}$ of charged hadrons around mid-$\eta $ as shown in
Figs. \ref{v13eta} (b) and \ref{vnPTeta12}.

\section{$p_{T}$-dependence of anisotropic flows at large rapidity}

\label{largeeta}

\begin{figure}[th]
\includegraphics[width=3.5in,height=2.5in]{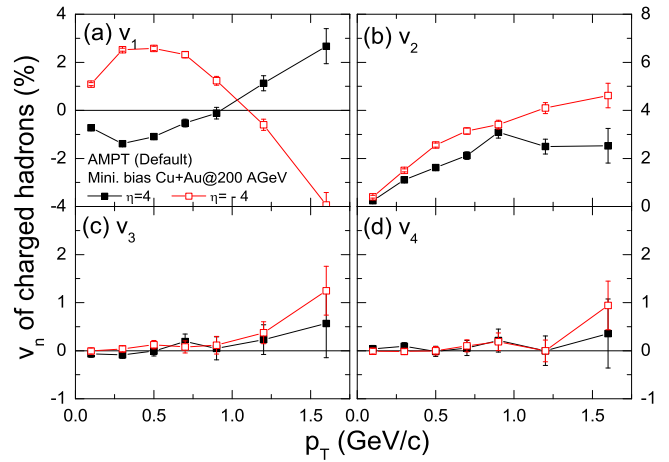}
\caption{{\protect\small (Color online) }$p_{T}${\protect\small -dependence
of }$v_{1}${\protect\small \ (a), }$v_{2}${\protect\small \ (b), }$v_{3}$%
{\protect\small \ (c) and }$v_{4}${\protect\small \ (d) for charged hadrons
at large forward pseudorapidity ($\protect\eta =4$, }{\protect\small solid
squares) and backward pseudorapidity ($\protect\eta =-4$, }{\protect\small %
open squares) from minimum bias events of Cu+Au collisions at }$\protect%
\sqrt{s}=200${\protect\small \ AGeV.}}
\label{vnPTBFdef}
\end{figure}

For hadrons at large rapidities ($\left\vert \eta \right\vert \geqslant 3$)
in heavy ion collisions at \textrm{RHIC}, the initial dynamics is expected
to be dominated by the string degree of freedom as they are produced later
in time when the volume of the system is large and the energy density is
low. Indeed, it has been shown that the default \textrm{AMPT} model in the
scenario without string melting can describe simultaneously data on $v_{1}$
and $v_{2}$ at large pseudorapidity ($\left\vert \eta \right\vert \geqslant
3 $) in Au+Au collisions at $\sqrt{s}=200$ AGeV \cite{v13plb05}. Therefore,
it is of interest to study anisotropic flows at large rapidities in
asymmetric Cu+Au collisions at \textrm{RHIC} based on the \textrm{AMPT}
model without string melting. In Fig. \ref{vnPTBFdef}, we show the predicted 
$p_{T}$-dependence of $v_{1}$, $v_{2}$, $v_{3}$, and $v_{4}$ of charged
hadrons at forward ($\eta =4$, solid squares) and backward ($\eta =-4$, open
squares) pseudorapidities from minimum bias events of Cu+Au collisions at $%
\sqrt{s}=200$ \textrm{AGeV}. It is seen that the directed flow $v_{1}(p_{T})$
shown in Fig. \ref{vnPTBFdef} (a) exhibits a strong asymmetry in forward and
backward rapidities, with charged hadrons at backward pseudorapidity having
a much stronger $v_{1}(p_{T})$ than that at forward pseudorapidity.
Furthermore, $v_{1}$ at forward pseudorapidity changes from negative to
positive values at a balance transverse momentum of about $0.8$ \textrm{GeV}/%
\textsl{c} while $v_{1}$ at backward pseudorapidity changes from positive to
negative values at a balance transverse momentum of about $1.2$ \textrm{GeV}/%
\textsl{c}. A balance transverse momentum for $v_{1}$ has also been observed
in symmetric Au+Au collisions as shown in Fig. 4 of Ref.\cite{v13plb05}, and
its existence has been attributed to the presence of transverse radial
expansion \cite{voloshin97}.

A clear asymmetry of the elliptic flow $v_{2}$ in forward and backward
rapidities is also seen in Fig. \ref{vnPTBFdef} (b), with charged hadrons at
backward pseudorapidity showing a stronger $v_{2}(p_{T})$ than that at
forward pseudorapidity. A similar asymmetry in forward and backward
rapidities is observed in the higher-order anisotropic flow $v_{3}(p_{T})$
at higher $p_{T}$ as shown in Fig. \ref{vnPTBFdef} (c). The higher-order
anisotropic flow $v_{4}(p_{T})$ shown in Fig. \ref{vnPTBFdef} (d) is
essentially zero and it is not clear if it has an asymmetry in forward and
backward rapidities.

\begin{figure}[th]
\includegraphics[scale=0.9]{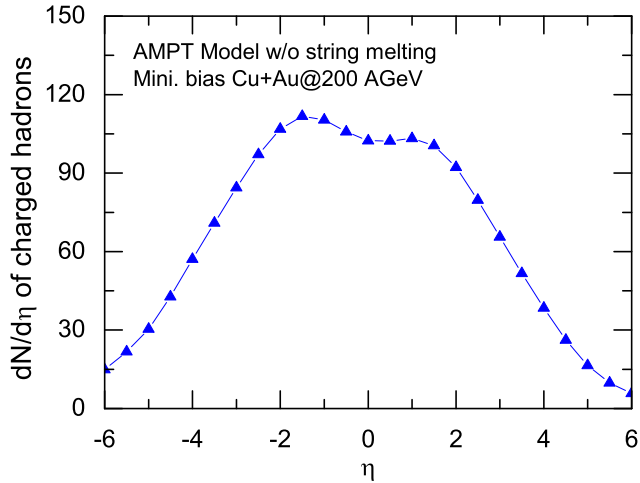}
\caption{{\protect\small (Color online) Charged hadron pseudorapidity
density for minimum bias events of Cu+Au collisions at }$\protect\sqrt{s}%
=200 ${\protect\small \ AGeV from the AMPT model in the scenario without
string melting.}}
\label{dNdETAchg}
\end{figure}

The observed strong asymmetry of charged hadron directed flow $v_{1}$ and
elliptic flow $v_{2}$ in large forward and backward pseudorapidities may be
related to the charged hadron pseudorapidity density, which is shown in Fig. %
\ref{dNdETAchg} for minimum bias events of Cu+Au collisions at $\sqrt{s}=200$
\textrm{AGeV} from the \textrm{AMPT} model in the scenario without string
melting.{\small \ }It is seen that the charged hadron pseudorapidity density
is asymmetric in forward and backward rapidities. Its value at backward
pseudorapidity (Au-like rapidity) is larger than that at forward
pseudorapidity (Cu-like rapidity), leading then to more hadronic
rescattering and thus stronger $v_{1}$\ and $v_{2}$ at backward
pseudorapidity as observed above.

\section{Summary}

\label{summary}

Using the \textrm{AMPT} model that includes both initial partonic and final
hadronic interactions, we have studied the anisotropic flows $v_{1}$, $v_{2}$%
, $v_{3}$, and $v_{4}$ of charged hadrons in asymmetric Cu+Au collisions at 
\textrm{RHIC}. Comparing with AMPT results from symmetric Au+Au collisions
shown in Refs.\cite{chen04,v13plb05}, we find that charged hadrons produced
around midrapidity in asymmetric Cu+Au collisions display a stronger
directed flow $v_{1}$ and their elliptic flow is also more sensitive to the
parton cross section used in the parton cascade. Furthermore, while
higher-order $v_{3}$ and $v_{4}$ are small at all rapidities, both $v_{1}$
and $v_{2}$ in these collisions are appreciable and show an asymmetry in the
forward and backward rapidities. This asymmetry is present even around
midrapidity for $v_{1}$. Experimental verification of these predictions for
asymmetric heavy-ion collisions at \textrm{RHIC} will be very useful in
understanding the dynamics of asymmetric collisions. In addition,
experimentally identifing the projectile and target sides of the reaction
plane in symmetric collisions is rather difficult. This has led to a
considerable delay in identifying the odd flow components, and the accuracy
of the measurements of these harmonics is still behind the accuracy and
detail achieved in the studies of the elliptic flow. Study of collisions
with asymmetry systems can contribute to a significant improvement of the
directed flow studies.

\begin{acknowledgments}
This work was supported in part by the National Natural Science Foundation
of China under Grant Nos. 10105008 and 10575071 (LWC) as wll as by the US
National Science Foundation under Grant No. PHY-0457265 and the Welch
Foundation under Grant No. A-1358 (CMK).
\end{acknowledgments}

\end{document}